\documentclass[preprint,showpacs,amsmath,amssymb,superscriptaddress]{revtex4}
\usepackage{graphicx}% Include figure files
\usepackage{dcolumn}% Align table columns on decimal point
\usepackage{bm}% bold math
\newcommand{\nc}{\newcommand}
\nc{\rnc}{\renewcommand}
\nc{\nn}{\nonumber}
\nc{\ch}{\cosh}
\nc{\sh}{\sinh}
\rnc{\th}{\tanh}
\nc{\db}{\displaybreak[0]\\}
\nc{\bra}{\langle}
\nc{\ket}{\rangle}
\rnc{\(}{\left(}
\rnc{\)}{\right)}
\nc{\xxx}{$XXX \,$}
\nc{\lam}{\lambda}
\nc{\lams}{\lam_1,\lam_2|\ldots|\lam_{2l-1},\lam_{2l}||\lam_{2l+1},\ldots,\lam_n}
\nc{\lamsd}{\lam_1,\lam_2|\ldots|\lam_{2l-1},\lam_{2l}||\lam_{2l+1},\ldots,\lam_{n-1}}
\nc{\Ak}{A_{n,l}^{\kappa}}
\nc{\Qk}{Q_{n,l}^{\kappa}}
\begin{document}
%
%%%%%%%%%%%%%%%%%%%%%%%%%%%%%%%%%%%%%%%%%%%%%%%%%%%%%%%%%%%%%%%%%%%%%%
\title{Fifth-neighbor spin-spin correlator for the anti-ferromagnetic Heisenberg chain}
%%%%%%%%%%%%%%%%%%%%%%%%%%%%%%%%%%%%%%%%%%%%%%%%%%%%%%%%%%%%%%%%%%%%%%
%
%%%%%%%%%%%%%%%%%%%%%%%%%%%%%%%%%%%%%%%%%%%%%%%%%%%%%%%%%%%%%%%%%%%%%%
\author{Jun Sato}
\email[]{junji@issp.u-tokyo.ac.jp}
\affiliation{Institute for Solid State Physics, University of Tokyo, 
Kashiwanoha 5-1-5, Kashiwa, Chiba 277-8571, Japan}

\author{Masahiro Shiroishi}
\email[]{siroisi@issp.u-tokyo.ac.jp}
\affiliation{Institute for Solid State Physics, University of Tokyo, 
Kashiwanoha 5-1-5, Kashiwa, Chiba 277-8571, Japan}

%\author{Minoru Takahashi}
%\email[]{mtaka@issp.u-tokyo.ac.jp}
%\affiliation{Institute for Solid State Physics, University of Tokyo, 
%Kashiwanoha 5-1-5, Kashiwa, Chiba 277-8571, Japan}

\date{\today}
%%%%%%%%%%%%%%%%%%%%%%%%%%%%%%%%%%%%%%%%%%%%%%%%%%%%%%%%%%%%%%%%%%
%
%%%%%%%%%%%%%%%%%%%%%%%%%%%%%%%%%%%%%%%%%%%%%%%%%%%%%%%%%%%%%%%%%%
%Abstract
%%%%%%%%%%%%%%%%%%%%%%%%%%%%%%%%%%%%%%%%%%%%%%%%%%%%%%%%%%%%%%%%%%

\begin{abstract}
We study the generating function of the spin-spin  correlation functions 
in the ground state of the anti-ferromagnetic spin-1/2 Heisenberg chain 
without magnetic field. We have found its fundamental functional relations 
from those for general correlation functions, which originate in the quantum 
Knizhnik-Zamolodchikov equation. Using these relations, we have calculated 
the explicit form of the generating functions up to ${n=6}$. Accordingly 
we could obtain the spin-spin correlator ${\langle S_j^z S_{j+k}^z \rangle}$ 
up to ${k=5}$.
\end{abstract}

\maketitle

%%%%%%%%%%%%%%%%%%%%%%%%%%%%%%%%%%%%%%%%%%%%%%%%%%%%%%%%%%%%%%%%%%%
%
%%%%%%%%%%%%%%%%%%%%%%%%%%%%%%%%%%%%%%%%%%%%%%%%%%%%%%%%%%%%%%%%%%
%Main Text
%%%%%%%%%%%%%%%%%%%%%%%%%%%%%%%%%%%%%%%%%%%%%%%%%%%%%%%%%%%%%%%%%%

Recently there have been rapid developments in the investigation of 
the exact calculation of the correlation functions for the spin-1/2  
Heisenberg chains. Especially for the ground state without 
magnetic field, explicit analytical form of several correlation 
functions have been calculated in the thermodynamic limit. In this 
letter we shall report further results for exact calculation of  
correlation functions for the antiferromagnetic Heisenberg  ${XXX}$ chain, 
whose Hamiltonian is given by
\begin{align}
\mathcal{H} = \sum_{j= -\infty}^{\infty} 
\left[ S_{j}^x S_{j+1}^x + S_{j}^y S_{j+1}^y + S_{j}^z S_{j+1}^z  \right],
\label{Hamiltonian}
\end{align}
where ${S_j^{\alpha} = \sigma_j^{\alpha}/2}$ with  ${\sigma_j^{\alpha}}$ being the 
Pauli matrices acting on the ${j}$-th site. The Hamiltonian (\ref{Hamiltonian}) 
can be diagonalized by Bethe ansatz and many bulk physical quantities have been 
evaluated in the thermodynamic limit \cite{Bethe31, TakaBook}. On the other hand, 
the exact calculation of the correlation functions is still an ongoing problem.  

First, let us list the known exact results of the two-point spin-spin 
correlators ${\langle S_j^z S_{j+k}^z \rangle}$, which are physically most important 
correlation functions : 
\begin{align}
\left\langle S_j^z S_{j+1}^z \right\rangle 
%&= \frac{1}{12} - \frac{1}{3}\ln 2
&= \frac{1}{12} - \frac{1}{3}\zeta_a(1)
 = -0.147715726853315\cdots, \label{nearest_neighbor} \\
\left\langle S_j^z S_{j+2}^z \right\rangle 
%&= \frac{1}{12} - \frac{4}{3} \ln 2 +\frac{3}{4} \zeta(3) 
&=\frac{1}{12} - \frac{4}{3} \zeta_a(1) + \zeta_a(3)
 =  0.060679769956435\cdots, \label{second_neighbor} \\
\left\langle S_{j}^{z} S_{j+3}^{z} \right\rangle 
& = \frac{1}{12} - 3 \zeta_a(1) + \frac{74}{9} \zeta_a(3) - \frac{56}{9} \zeta_a(1) \zeta_a(3) 
- \frac{8}{3} \zeta_a(3)^2  \nonumber  \\
& - \frac{50}{9} \zeta_a(5) + \frac{80}{9} \zeta_a(1) \zeta_a(5) 
\nonumber  \\
&= -0.050248627257235\cdots, \label{third_neighbor} \\
\left\langle S_{j}^{z} S_{j+4}^{z} \right\rangle
& = \frac{1}{12} - \frac{16}{3} \zeta_a(1)  + \frac{290}{9} \zeta_a(3) - 72 \zeta_a(1) \zeta_a(3) 
- \frac{1172}{9} \zeta_a(3)^2  - \frac{700}{9} \zeta_a(5)  \nonumber  \\
& 
+ \frac{4640}{9} \zeta_a(1) \zeta_a(5) - \frac{220}{9} \zeta_a(3) \zeta_a(5) - \frac{400}{3} \zeta_a(5)^2 
\nonumber  \\
& + \frac{455}{9} \zeta_a(7) - \frac{3920}{9} \zeta_a(1) \zeta_a(7) 
+ 280 \zeta_a(3) \zeta_a(7)  \nonumber \\
&=  0.034652776982728\cdots. \label{fourth_neighbor}   
\end{align}
Here ${\zeta_a(s)}$ is the alternating zeta function defined by 
${\zeta_a(s)  \equiv \sum_{n=1}^{\infty} (-1)^{n-1}/n^s}$, which is related to the 
Riemann zeta function ${\zeta(s)  = \zeta_a(s)/(1-2^{1-s})}$.
Note that the alternating zeta function is regular at ${s=1}$ and is given by ${\zeta_a(1)}=\ln 2$.
Thus analytical form of the spin-spin correlators ${\langle S_j^z S_{j+k}^z \rangle}$ have been 
obtained up to ${k=4}$ so far. The nearest-neighbor correlator (\ref{nearest_neighbor}) was 
obtained from Hulth\'{e}n's result of the ground state energy in 1938 \cite{Hulthen38}. 
The second-neighbor correlator (\ref{second_neighbor}) was obtained by Takahashi in 1977 
\cite{Takahashi77} via the strong coupling expansion of the ground state energy for the half-filled Hubbard 
chain. After that it had taken some long time before the explicit form of the 
third-neighbor correlator (\ref{third_neighbor}) was obtained 
by Sakai, Shiroishi, Nishiyama and Takahashi in 2003 \cite{Sakai03}. 
They applied the Boos-Korepin method \cite{Boos01,Boos01n2} to evaluate the multiple integral 
formula for correlation functions developed by Kyoto group \cite{Jimbo92,Korepin94,Nakayashiki94,
JMBook,Jimbo96}(see also \cite{Kitanine00}). The Boos-Korepin method was originally devised to 
calculate a special correlation function called the emptiness formation probability (EFP) \cite{Korepin94}, which is defined by 
\begin{align}
P(n) \equiv \left\langle \prod_{j=1}^{n} \left( \frac{1}{2} +S_j^z \right) \right\rangle.
\end{align}
This is the probability to find a ferromagnetic string of length ${n}$ in the ground state. 
The multiple integral formula for ${P(n)}$ as well as other correlation functions among the 
adjacent ${n}$-spins consists of the ${n}$-dimensional integrals.  

By the use of Boos-Korepin method, we can reduce the multiple integrals to 
one-dimensional ones by transforming the integrand to a certain canonical 
form. In the case of the present ${XXX}$ model, the remaining 
one-dimensional integrals can be further expressed in terms of ${\zeta_a(s), s \in {\mathbb Z}_{\ge 1}}$. 
Hence, {\it in principle}, we can obtain the explicit form of arbitrary correlation functions 
by Boos-Korepin method. However, ${\it practically}$ it becomes a tremendously hard task to 
calculate the canonical form as the integral dimension increases. Actually ${P(5)}$ is the only 
correlation function which was calculated by Boos-Korepin method among those for five lattice sites 
 \cite{Boos02}. 

To overcome such a difficulty, Boos, Korepin and Smirnov invented an alternative method to obtain 
${P(n)}$ \cite{Boos03}. They consider the inhomogeneous ${XXX}$ model, where each site carries an 
inhomogeneous parameter ${\lam_j}$. Accordingly the correlation functions depend on the parameters 
${\lam_j}$, for example, ${P_n(\lam_1,...,\lam_n)}$. Boos, Korepin and Smirnov derived simple functional 
relations, which ${P_n(\lam_1,...,\lam_n)}$ should satisfy from the underlying quantum 
Knizhnik-Zamolodchikov (qKZ) equations. Together with a simple ansatz 
for the final form of the ${P_n(\lam_1,...,\lam_n)}$, they have shown the problem reduces to solving large linear systems of equations for rational coefficients. In this way they could calculate 
${P_n(\lam_1,...,\lam_n)}$ up to ${n=6}$. Especially, by taking the homogeneous limit ${\lam_j \to 0}$, they 
obtained the analytical form of ${P(n)}$ up to ${n=6}$. Thus this new method is more powerful 
than the original Boos-Korepin method to evaluate the multiple integrals. It was recently 
generalized to calculate arbitrary inhomogeneous correlation functions by Boos, Shiroishi and Takahashi \cite{Boos05}. 
They have calculated all the independent correlation functions among  five lattice sites and especially the 
fourth-neighbor correlator (\ref{fourth_neighbor}). In this letter we explore these functional methods 
further.  Especially main new aspect is to introduce the generating function 
of the spin-spin correlators \cite{Izergin85,Essler96,Kitanine02,Kitanine05,Kitanine05n2} into the scheme, which allows us to calculate the 
${\langle S_j^z S_{j+k}^z \rangle}$ more efficiently. 

We remark that Boos, Jimbo, Miwa, Smirnov and Takeyama have obtained more explicit recursion 
relations for arbitrary correlation functions \cite{BJMST1} (see also \cite{BJMST2}) by 
investigating the qKZ equations more profoundly. Their formulas have proven the ansatz for the final form of 
the inhomogeneous correlation functions.  However, the formulas contain unusual trace functions with 
a {\it fractional} dimension, effective evaluation of which has not been developed yet.

Here we define the generating function by 
\begin{align}
P_n^{\kappa}(\lam_1,\lam_2,...,\lam_n) & \equiv 
\left\langle \prod_{j=1}^n 
\left\{ \left(\frac{1}{2} + S_j^z \right) 
+ \kappa \left( \frac{1}{2} - S_j^z \right) \right\} \right\rangle (\lam_1,\lam_2,...,\lam_n), 
\end{align}
where ${\kappa}$ is a parameter. Once we obtain the generating functions, we can calculate the two-point 
spin-spin correlators through the relation \cite{Kitanine05}
\begin{align}
\left\langle S_{1}^{z} S_{k+1}^{z} \right\rangle (\lam_1,...,\lam_{k+1})
& = \frac{1}{2} \frac{\partial^2}{\partial \kappa^2} 
\Big\{ P_{k+1}^{\kappa}(\lam_1,\lam_2,...,\lam_{k+1}) 
-P_{k}^{\kappa}(\lam_1,...,\lam_{k}) \nonumber \\
& - P_{k}^{\kappa}(\lam_2,...,\lam_{k+1})
+P_{k-1}^{\kappa}(\lam_2,...,\lam_{k}) \Big\} \Bigg|_{\kappa
=1} - \frac{1}{4}, \nonumber  \\
\left\langle S_{1}^{z} S_{k+1}^{z} \right\rangle &=\lim_{\lam_j \to 0} 
\left\langle S_{1}^{z} S_{k+1}^{z} \right\rangle (\lam_1,...,\lam_{k+1}). 
\label{relation}
\end{align}

Note that ${P_n^{\kappa}(\lam_1,\lam_2,...,\lam_n)}$ is a natural generalization of the EFP as 
\begin{align}
& P_n^{\kappa=0}(\lam_1,\lam_2,...,\lam_n)
= P_n(\lam_1,\lam_2,...,\lam_n),
\end{align}
with other  relations 
\begin{align}
& P_n^{\kappa=1}(\lam_1,\lam_2,...,\lam_n)= 1, \\
& P_n^{\kappa=-1}(\lam_1,\lam_2,...,\lam_n)= 
2^n \langle \prod_{j=1}^{n} S_j^z \rangle (\lam_1,\lam_2,...,\lam_n). 
\end{align}
From the qKZ equations for the correlation functions \cite{Boos05,BJMST1,BJMST2}, we find 
the generating function satisfies the following functional relations : 
\begin{itemize}

\item
Translational invariance

\begin{align}
P_n^{\kappa}(\lam_1+x,\ldots,\lam_n+x)=
P_n^{\kappa}(\lam_1,\ldots,\lam_n)
\label{translation}
\end{align}

\item
Negating relation

\begin{align}
P_n^{\kappa}(\lam_1,\ldots,\lam_n)=
P_n^{\kappa}(-\lam_1,\ldots,-\lam_n)
\label{negating}
\end{align}

\item
Symmetry relation
\begin{align}
P_n^{\kappa}(\lam_1,\ldots,\lam_j,\lam_{j+1},\ldots,\lam_n)=
P_n^{\kappa}(\lam_1,\ldots,\lam_{j+1},\lam_{j}, \ldots,\lam_n)
\label{symmetry}
\end{align}

\item
First recurrent relation

\begin{align}
P_n^{\kappa}(\lam_1,\ldots,\lam_{n-1},\lam_{n-1} \pm 1)=
\kappa P_{n-2}^{\kappa}(\lam_1,\ldots,\lam_{n-2})
\label{recurrent1}
\end{align}

\item
Second recurrent relation

\begin{align}
\lim_{\lam_n \rightarrow i\infty}
P_n^{\kappa}(\lam_1,\ldots,\lam_{n-1},\lam_{n})=
\frac{1 +\kappa}{2} P_{n-1}^{\kappa}(\lam_1,\ldots,\lam_{n-1})
\label{recurrent2}
\end{align}

\end{itemize}
Here and hereafter we follow the notations in \cite{BJMST1} for the spectral 
parameter ${\lambda_j}$ and also the transcendental function ${\omega(\lambda)}$ below. 
One easily sees, these functional relations reduce to those for the EFP,  
$P_n(\lam_1,\ldots, \lam_{n})$ \cite{Boos03} if we set ${\kappa=0}$. 
According to the (proved) general form of the correlation functions and also by the symmetry 
relation (\ref{symmetry}), we can assume the generating function in the form  
\begin{align}
&P_n^{\kappa}(\lam_1,\ldots,\lam_n)= \nonumber \\
&
\sum_{l=0}^{[\frac{n}{2} ]}
\left\{ A_{n,l}^{\kappa}(\lams)
\prod_{j=1}^{l} \omega(\lam_{2j-1}-\lam_{2 j}) 
+ {\rm permutations} \right\},
\label{ansatz}
\end{align}
where
\begin{align}
\omega(\lam)=\sum_{k=1}^{\infty}(-1)^k 
\frac{2 k (\lam^2-1)}{\lam^2-k^2} + \frac{1}{2},
\label{omega}
\end{align}
is a generating function of the alternating zeta values
\begin{align}
\omega(\lam)=2 (\lam^2-1) \sum_{k=0}^{\infty} \lam^{2 k}
\zeta_a(2 k+1) + \frac{1}{2}.
\end{align}
Moreover the function $\omega(\lam)$ has the following properties:
\begin{align}
\omega(i\infty)=0, \ \ \ \ \ \omega(\lam\pm1)=\alpha(\lam)+\gamma_{\pm}(\lam)\omega(\lam).
\label{omegaRelation}
\end{align}
where
\begin{align}
\alpha(\lam)=\frac{3}{2}\frac{1}{\lam^2-1}, \ \ \ \ \ \
\gamma_{\pm}(\lam)=-\frac{\lam(\lam\pm2)}{\lam^2-1}.
\label{AlphaGamma}
\end{align}
Note also that 
$A_{n,l}^{\kappa}(\lams)$ are rational functions 
depending on the parameter ${\kappa}$ with known denominators:
\begin{align}
&A_{n,l}^{\kappa}(\lams) \nonumber  \\
&=
\frac{
\prod_{k=1}^l\lam_{2k-1,2k}
\prod_{2l+1\leq k<j\leq n}\lam_{kj}
}
{
\prod_{1\leq k<j\leq n}\lam_{kj}
}
\,Q_{n,l}^{\kappa}(\lams),
\label{q}
\end{align} 
where $\lam_{kj}=\lam_k-\lam_j$ and ${Q_{n,l}^{\kappa}(\lams)}$ are some polynomials.

Our main proposal is to determine ${A_{n,l}^{\kappa}(\lams)}$ 
and therefore the explicit form of the 
generating functions from the functional relations (\ref{translation})--(\ref{recurrent2}) 
together with the ansatz (\ref{ansatz}).
%  We omit here the details of calculations, which are 
%actually similar to those for the EFP in \cite{Boos03}.  
Simple calculations yield that the second recurrent relation (\ref{recurrent2})
is equivalent to the following recursion equation for $\Ak$:
\begin{align}
\lim_{\lam_n \rightarrow i\infty}
\Ak(\lams)=
\frac{1 +\kappa}{2} 
A_{n-1,l}^{\kappa}
(\lamsd).
\label{Arecurrent2}
\end{align}
Note that it is easy to see 
\begin{align}
A_{n,0}^{\kappa}(||\lam_1,\ldots,\lam_n)= \left(\frac{1+\kappa}{2}\right)^n.
\end{align}
for any ${n}$.
The first recurrent relation (\ref{recurrent1}) 
can not be reduced to a general formula for $\Ak$.
Instead, we write down the equations for $\Ak$ corresponding to
the first recurrent relation up to $n=3$:
\begin{align}
&
\(\frac{1+\kappa}{2}\)^2-\frac{3}{2}A_{2,1}^{\kappa}(\lam_1,\lam_1\pm1||)=\kappa,\\
&
\(\frac{1+\kappa}{2}\)^3
+\alpha(\lam_{12})A_{3,1}^{\kappa}(\lam_1,\lam_2\pm1||\lam_2)
-\frac{3}{2}A_{3,1}^{\kappa}(\lam_2,\lam_2\pm1||\lam_1)
=\kappa\(\frac{1+\kappa}{2}\),\\
&
A_{3,1}^{\kappa}(\lam_1,\lam_2||\lam_2\pm1)
+\gamma_{\mp}(\lam_{12})A_{3,1}^{\kappa}(\lam_1,\lam_2\pm1||\lam_2)
=0.
\label{Arecurrent1}
\end{align}
Solving these equations,
we have calculated the explicit forms of ${Q_{n,l}^{\kappa}}$ up to 
${n=5}$, which are given by 
\begin{align}
& Q_{2,1}^{\kappa}(\lam_1,\lam_2||) = \frac{1}{6}(1-\kappa)^2, \ \ \ \ 
Q_{3,1}^{\kappa}(\lam_1,\lam_2||\lam_3) = 
\frac{1+\kappa}{2}(1+\lam_{13} \lam_{23})Q_{2,1}^{\kappa}(\lam_1,\lam_2||)
, \nonumber \\
& Q_{4,1}^{\kappa}(\lam_1,\lam_2||\lam_3,\lam_4) 
= 
\frac{1+\kappa}{2}(1+\lam_{14} \lam_{24})Q_{3,1}^{\kappa}(\lam_1,\lam_2||\lam_3)
 -\frac{(1-\kappa)^4}{120} 
(\lam_{12}^2-4), \nonumber \\
& Q_{4,2}^{\kappa}(\lam_1,\lam_2|\lam_3,\lam_4||) 
= 
\frac{1}{24} \left( 1+\frac{\kappa}{3} \right)(1+3 \kappa)(1-\kappa)^2 
+\frac{(1-\kappa)^4}{90} \left( \lam_{12}^2 -\frac{3}{2} \right)
\left( \lam_{34}^2 -\frac{3}{2} \right) \nonumber  \\
& \ \ \ \ \ \ \ \ \ \ 
+\frac{(1+ \kappa )^2(1-\kappa)^2 }{36}(1+\lam_{13}\lam_{24})(1+\lam_{14}\lam_{23}) 
- \frac{\kappa(1-\kappa)^2 }{18} (1-\lam_{13}\lam_{24})(1-\lam_{14}\lam_{23}) 
, \nonumber \\
& Q_{5,1}^{\kappa}(\lam_1,\lam_2||\lam_3,\lam_4,\lam_5) 
=\frac{1+\kappa}{2}(1+\lam_{15} \lam_{25})
Q_{4,1}^{\kappa}(\lam_1,\lam_2||\lam_3,\lam_4) \nonumber  \\
& \hspace{5cm}
-\frac{1+\kappa}{2}\frac{(1-\kappa)^4}{120}
(\lam_{12}^2-4)(4+\lam_{13}\lam_{23}+\lam_{14}\lam_{24}),
 \nonumber  \\
& Q_{5,2}^{\kappa}(\lam_1,\lam_2|\lam_3,\lam_4||\lam_5) 
=\frac{1+\kappa}{2}(1+\lam_{15}\lam_{25})(1+\lam_{35}\lam_{45})
Q_{4,2}^{\kappa}(\lam_1,\lam_2|\lam_3,\lam_4||) \nonumber  \\
&
+\frac{(1-\kappa)^2}{360}(\lam_{12}^2-4)(\lam_{34}^2-4)
\left\{
(1+\kappa)(1-\kappa)^2(\lam_{14}\lam_{23}+\lam_{13}\lam_{24})
+5(1+\kappa+\kappa^2+\kappa^3)
\right\} \nonumber  \\
&
-\frac{\kappa(1+\kappa)(1-\kappa)^2}{36}
\left\{
10-2(\lam_{12}^2+\lam_{34}^2+\lam_{14}\lam_{23}\lam_{13}\lam_{24})+(\lam_{14}\lam_{23}+\lam_{13}\lam_{24}+2)(\lam_{15}\lam_{25}+\lam_{35}\lam_{45}-1)
\right\}.
 \nonumber \\
\end{align} 
One can confirm, the previous known results of two-point spin-spin correlators (\ref{nearest_neighbor}),...,(\ref{fourth_neighbor}) 
are reproduced through the relation (\ref{relation}). 
In a similar way, we have calculated ${P_6^{\kappa}(\lam_1,\ldots,\lam_{6})}$. 
Since its expression is too complicated, we present here only the final two new 
results of correlation 
functions among six lattice sites.  The first one is the two-point fifth-neighbor spin-spin correlator,   
\begin{align}
& \left\langle S_{j}^{z} S_{j+5}^{z} \right\rangle \nonumber  \\
& = \frac{1}{12} - \frac{25}{3} \zeta_a(1) + \frac{800}{9} \zeta_a(3)- \frac{1192}{3} \zeta_a(1) \zeta_a(3) 
  - \frac{15368}{9} \zeta_a(3)^2 - 608 \zeta_a(3)^3 - \frac{4228}{9} \zeta_a(5) \nonumber  \\
& + \frac{64256}{9} \zeta_a(1) \zeta_a(5) 
 - \frac{976}{9} \zeta_a(3) \zeta_a(5) + 3648 \zeta_a(1) \zeta_a(3) \zeta_a(5) 
 - \frac{3328}{3} \zeta_a(3)^2 \zeta_a(5) - \frac{76640}{3} \zeta_a(5)^2 \nonumber  \\
 &
 + \frac{66560}{3} \zeta_a(1) \zeta_a(5)^2 + \frac{12640}{3} \zeta_a(3) \zeta_a(5)^2 
 + \frac{6400}{3} \zeta_a(5)^3 + \frac{9674}{9} \zeta_a(7)  - \frac{225848}{9} \zeta_a(1) \zeta_a(7) \nonumber  \\
& + 56952 \zeta_a(3) \zeta_a(7) 
 - \frac{116480}{3} \zeta_a(1) \zeta_a(3) \zeta_a(7) - \frac{35392}{3} \zeta_a(3)^2 \zeta_a(7) 
 + 7840 \zeta_a(5) \zeta_a(7) \nonumber  \\
& -8960 \zeta_a(3) \zeta_a(5) \zeta_a(7) - \frac{66640}{3} \zeta_a(7)^2 
 + 31360 \zeta_a(1) \zeta_a(7)^2 - 686 \zeta_a(9) + 18368 \zeta_a(1) \zeta_a(9) \nonumber  \\ 
& - 53312 \zeta_a(3) \zeta_a(9) + 35392 \zeta_a(1) \zeta_a(3) \zeta_a(9) 
 + 16128 \zeta_a(3)^2 \zeta_a(9) \nonumber  \\
& + 38080 \zeta_a(5) \zeta_a(9)  - 53760 \zeta_a(1) \zeta_a(5) \zeta_a(9) \nonumber  \\
& = -0.030890366647609\cdots. 
\label{fifth-neighbor}
\end{align}
Another one is a six-spin correlation function
\begin{align}
& \lim_{\lambda_{j} \to 0} P_6^{\kappa=-1}(\lambda_1,\ldots,\lambda_6)
= 2^6 \ \langle \prod_{j=1}^6 S_j^z  \rangle \nonumber  \\
& = \frac{1}{7} - 12 \zeta_a(1) + \frac{1112}{5} \zeta_a(3)- \frac{3776}{3} \zeta_a(1) \zeta_a(3) 
  - \frac{100736}{15} \zeta_a(3)^2 - \frac{352768}{135} \zeta_a(3)^3 - \frac{71656}{35} \zeta_a(5) \nonumber  \\
& + \frac{442496}{15} \zeta_a(1) \zeta_a(5) + \frac{15104}{15} \zeta_a(3) \zeta_a(5) + \frac{705536}{45} 
\zeta_a(1) \zeta_a(3) \zeta_a(5) - \frac{212992}{45} \ \zeta_a(3)^2 \zeta_a(5) 
 \nonumber  \\
 &
 - \frac{6796736}{63} \zeta_a(5)^2 + \frac{851968}{9} \zeta_a(1) \zeta_a(5)^2 + \frac{161792}{9} \zeta_a(3) \zeta_a(5)^2 
 + \frac{1723520}{189} \zeta_a(5)^3 + \frac{32432}{5} \zeta_a(7) \nonumber \\ 
& - \frac{350336}{3} \zeta_a(1) \zeta_a(7) + 241888 \zeta_a(3) \zeta_a(7) - \frac{1490944}{9} \zeta_a(1) \zeta_a(3) 
\zeta_a(7) 
- \frac{2265088}{45} \zeta_a(3)^2 \zeta_a(7) \nonumber  \\
&  + \frac{312064}{9} \zeta_a(5) \zeta_a(7) -\frac{344704}{9} \zeta_a(3) \zeta_a(5) \zeta_a(7) - \frac{833168}{9} \zeta_a(7)^2 
 + \frac{1206464}{9} \zeta_a(1) \zeta_a(7)^2 
\nonumber  \\ 
& - \frac{23256}{5} \zeta_a(9) + \frac{443008}{5} \zeta_a(1) \zeta_a(9) - \frac{3437248}{15} \zeta_a(3) \zeta_a(9) + \frac{2265088}{15} \zeta_a(1) \zeta_a(3) \zeta_a(9) 
  \nonumber  \\
& + \frac{344704}{5} \zeta_a(3)^2 \zeta_a(9) + \frac{476096}{3} \zeta_a(5) \zeta_a(9)  - \frac{689408}{3} \zeta_a(1) \zeta_a(5) \zeta_a(9) \nonumber  \\
& = -0.440301669702626\cdots. 
\label{z6corr}
\end{align}
One can see that
these analytical expression have the same structure 
as $P(6)$ previously obtained in \cite{Boos03}.
Namely they are the polynomials of $\zeta_a(1),\ldots,\zeta_a(9)$ of degree $3$
with different rational coefficients.
%Probably other correlation functions among six lattice sites will have the same
%structure.
We have confirmed these results (\ref{fifth-neighbor}) and (\ref{z6corr}) by comparing with the 
numerical data by the exact diagonalization for the finite systems. 

In summary we have presented a new effective method to calculate the analytical form of the spin-spin 
correlators starting from the generating function. In particular we could obtain the fifth-neighbor correlator 
${\left\langle S_{j}^{z} S_{j+5}^{z} \right\rangle}$. The details of the calculations and the further 
results will be reported in a separate publication.  We are grateful to K.~Sakai, M.~Takahashi and Z.~Tsuboi for valuable discussions. 
This work is in part supported by Grant-in-Aid for the Scientific Research (B) No.~14340099 
from the Ministry of Education, Culture, Sports, Science and Technology, Japan. 
MS is supported by Grant-in-Aid for young scientists No.~14740228.

%%%%%%%%%%%%%%%%%%%%%%%%%%%%%%%%%%%%%%%%%%%%%%%%%%%%%%%%%%%%%%%%%%
%Bibliography
%%%%%%%%%%%%%%%%%%%%%%%%%%%%%%%%%%%%%%%%%%%%%%%%%%%%%%%%%%%%%%%%%%
%

\end{document}